\documentclass[twocolumn,showpacs,preprintnumbers,amsmath,amssymb,prb,superscriptaddress]{revtex4}

\usepackage{graphicx}
\usepackage{dcolumn}
\usepackage{bm}
\usepackage{epstopdf}
\usepackage{xcolor}

\begin{document}


\title{Real-space multiple scattering theory for superconductors with impurities}

\author{Tom G. Saunderson}
\email{t.saunderson@bristol.ac.uk}
\affiliation{HH Wills Physics Laboratory, University of Bristol, Tyndall Ave, BS8 1TL, United Kingdom}
\author{Zsolt Gy\H{o}rgyp\'al}%
\affiliation{Institute for Solid State Physics and Optics, Wigner Research Centre for Physics, Hungarian Academy of Sciences, P.O. Box 49, H-1525 Budapest, Hungary}
\author{James F. Annett}%
\affiliation{HH Wills Physics Laboratory, University of Bristol,
Tyndall Ave, BS8 1TL, United Kingdom}
\author{G\'abor Csire}%
\affiliation{Catalan Institute of Nanoscience and Nanotechnology (ICN2), CSIC, BIST, Campus UAB, Bellaterra, Barcelona, 08193, Spain}
\author{Bal\'azs \'Ujfalussy}%
\affiliation{Institute for Solid State Physics and Optics, Wigner Research Centre for Physics, Hungarian Academy of Sciences, P.O. Box 49, H-1525 Budapest, Hungary}
\author{Martin Gradhand}%
\affiliation{HH Wills Physics Laboratory, University of Bristol, Tyndall Ave, BS8 1TL, United Kingdom}

\date{\today}

\begin{abstract}
We implement the Bogoliubov-de~Gennes (BdG) equation in real-space using the screened Korringa-Kohn-Rostoker (KKR) method. This allows us to solve, self-consistently, the superconducting state for 3d crystals including substitutional impurities with a full normal-state DFT band structure. We apply the theoretical framework to bulk Nb with impurities. Without impurities, Nb has an anisotropic gap structure with two distinct peaks around the Fermi level. In the presence of non-magnetic impurities those peaks are broadened due to the scattering between the two bulk superconducting gaps, however the peaks remain separated. As a second example of self-consistent real-space solutions of the BdG equations we examine superconducting clusters embedded within a non-superconducting bulk metallic host. 
This allows us to estimate the coherence length of the superconductor and we show that, within our framework, the coherence length of the superconductor is related to the inverse of the gap size, just as in bulk BCS theory.
\end{abstract}

\pacs{Valid PACS appear here}
\maketitle


\section{\label{sec:1}Introduction}

Inhomogeneities in supperconductors have been of intense interest for many years. 
Impurities in bulk materials have been exploited to destroy superconductivity \cite{Li2015,Alloul2009}, create superconductivity  by doping \cite{Stacy1987,Wu1987,Enomoto1987}, determine the order parameter of superconductors \cite{VanDyke2016,Sprau2017,Kreisel2017,Kostin2018} and create bound Yu-Shiba-Rusinov states in superconducting materials \cite{Hatter2017,Ruby2018,Heinrich2018,Kezilebieke2018,Menard2019}. Furthermore, nanoscale structured superconducting materials have been engineered to provide artificial materials with desired characteristics, such as increased critical temperature T$_c$, or granular as well as percolative superconductivity.\cite{Deutscher2004}

Modelling inhomogeneous systems such as these generally requires real-space solutions of the Bogoliubov-de~Gennes (BdG) equations. Fetter~\cite{Fetter1965} was one of the first to use localised models to describe non-magnetic impurities in the superconducting state. It was shown that impurities in real materials will induce both spatial oscillations in the pairing potential $\Delta(r)$ and the electron density $\rho(r)$. In addition, the resonant enhancement of the scattering of quasiparticles with momentum near the Fermi momentum was identified. Later, Flatt\'e and Byers \cite{Flatte1997} provided insightful models into how magnetic and non-magnetic impurities behave in a free-electron s-wave superconducting medium. These models provided qualitative insight of a generic localised perturbation in a model superconductor, but lacked any quantitative predictive power to describe the complex impurity states which would occur in real materials. Materials specific information about the 
superconducting gap variation on the  Fermi surface were considered in realistic tight-binding models.~\cite{Wang2003,Pereg-Barnea2008,Zhang2009a,Hirschfeld2011,Hirschfeld2015,VanDyke2016,Kreisel2017,Sulangi2017} In all those cases the impurities were used as a probe to investigate the superconducting gap structure and order parameter for unconventional superconductors. This quasiparticle interference on real surfaces has been visualised experimentally using scanning tunnelling microscopy \cite{Hoffmann2002,McElroy2003,Fischer2007,Ji2008,Hanaguri2010,Allen2012,Allan2013,Zhou2013,Ronen2016,Avraham2018}  and provided powerful insight into the superconducting state. Since most of the theoretical approaches work in reciprocal space direct comparison to experiments will typically involve Fourier transformations of the direct real space analysis of the experiments. 

Understanding such inhomogenous systems at the \textit{ab initio} level poses significant challenges, even for conventional electron-phonon BCS driven superconductors. In the bulk, 
modelling of phonon mediated s-wave superconductors has been successful using modern DFT techniques \cite{Giustino2017}.  Incorporating impurities or nanoscale structured materials into these  \textit{ab initio}  methods would be possible in principle, but would become technically very challenging and computationally demanding. Even in bulk systems the full theory requires six-dimensional integrals both over the electron and phonon Brillouin zones, ${\bf k}$ and ${\bf Q}$. In systems without translational symmetry the corresponding real-space coupled electron-phonon equations would become significantly more difficult to solve. Similarly, for unconventional superconductors, models including impurities either use a 
simplified parametrization of the problem, or they are forced to use supercell approaches to incorporate the impurity site \cite{Choi2017}.




In this work we will exploit the Korringa-Kohn-Rostoker~(KKR) Greens function method which is
ideally suited to treat the real space impurity problem in a full quantitative \textit{ab initio }approach \cite{Ebert2011a}. We combine the first principles treatment of the impurity problem with the implementation of the Bogoliubov-de~Gennes (BdG) equations which we developed previously to describe superconductivity in periodic crystals and surfaces in k-space. \cite{Csire2015a,Saunderson2020} Within this framework a phenomenological parametrisation of the pairing interaction introduces the parameter $\Lambda$, which  is fixed by the experimentally found gap size. Such a treatment has been shown to reproduce experimentally observed gap anisotropies for various materials such as Nb, Pb and MgB$_2$~\cite{Saunderson2020}. It has also been used to develop a quantitative theory for triplet pairing in $LaNiC_2$ and $LaNiGa_2$.\cite{Csire2018a,Csire2020}

This method incorporates the full orbital character of real impurities in contrast to previous tight-binding models.
The explicit real space description will allow for more direct comparison to local experimental probes having direct access to the local density of states (LDOS). After a brief introduction on the specific implementation in Section \ref{sec:2}, the method will be tested with a range of impurities in Nb in order to explore the different effects of impurities showing distinct orbital character in Section~\ref{sec:3}. In Section~\ref{sec:4} we apply this method to granular superconductors, solving \textit{ab initio} a nanoscale 
superconducting cluster embedded in a normal metallic environment. We can view this system as a sort of inverse problem to that of an impurity in a bulk superconducting host.  
Solving the real-space BdG equations for the superconducting cluster allows us to make direct contact to the concept of the superconducting coherence length as applied to granular superconductivity. After Section~\ref{sec:4} we compare our calculations to one dimensional models to get a more fundamental understanding and a numerically easier access to the superconducting coherence length.

\section{\label{sec:2}Method}

This implementation will rely on the existing real-space screened KKR impurity code \cite{Gradhand_2010_imp} in combination with the BdG solver for the periodic lattices\cite{Saunderson2020}. Here, we will focus on the most relevant aspects crucially relevant for the consideration in real space impurity systems. All equations are given in Rydberg units.
The effective potentials within the theory of superconducting Density Functional Theory (DFT), exploiting the same approximations as highlighted in Ref.~\onlinecite{Saunderson2020}, are the electron potential $V_{eff}(\mathbf{r})$ and the effective pairing potential $\Delta_{eff}(\mathbf{r})$, 
\begin{align}
\label{eqn:Veff}
V_{eff}(\mathbf{r}) &= V_{ext}(\mathbf{r}) + \int d^3 r  \frac{\rho(\mathbf{r})}{|\mathbf{r}-\mathbf{r}'|} + \frac{\delta E_{xc}[\rho]}{\delta \rho(\mathbf{r})}, \\
\label{eqn:Deltaeff}
\Delta_{eff}(\mathbf{r}) &= \Lambda \chi(\mathbf{r}).
\end{align}
Here $\chi(\mathbf{r})$ is the anomalous density, $\Lambda$ is the interaction parameter and $E_{xc}[\rho]$ is the exchange correlation functional for the normal state. All densities are expressed via the Green's function
\begin{align}
\label{eqn:rho}
\rho(\mathbf{r}) = &-\frac{1}{\pi}\int^{\infty}_{-\infty} d\epsilon f(\epsilon) \mathrm{Im}\mathrm{Tr}G^{ee}(\epsilon,\mathbf{r},\mathbf{r}') \nonumber\\
&-\frac{1}{\pi}\int^{\infty}_{-\infty} d\epsilon [1-f(\epsilon)] \mathrm{Im}\mathrm{Tr}G^{hh}(\epsilon,\mathbf{r},\mathbf{r}'),\\
\label{eqn:chi}
\chi(\mathbf{r}) = &-\frac{1}{4\pi}\int^{\infty}_{-\infty} d\epsilon [1-2f(\epsilon)] \mathrm{Im} \mathrm{Tr} G^{eh} (\epsilon,\mathbf{r},\mathbf{r}') \nonumber \\
&-\frac{1}{4\pi}\int^{\infty}_{-\infty} d\epsilon [1-2f(\epsilon)]  \mathrm{Im} \mathrm{Tr} G^{he} (\epsilon,\mathbf{r},\mathbf{r}'),
\end{align}
where the Bogoliubov-de~Gennes Hamiltonian $\hat{H}_{BdG}(\mathbf{r})$ and Green's function $\hat{G}_{BdG}(z)$ are defined as
\begin{equation}
\hat{G}_{BdG}(z) =
\left(\begin{matrix}
\hat{G}^{ee}(z) & \hat{G}^{eh}(z) \\
\hat{G}^{he}(z) & \hat{G}^{hh}(z)
\end{matrix}\right) =
\big(z\hat{I} - \hat{H}_{BdG}\big)^{-1},
\end{equation}
with $\hat{H}_{BdG}(\mathbf{r}) = \langle \mathbf{r}|\hat{H}_{BdG}|\mathbf{r}\rangle$ and
\begin{align}
\hat{H}_{BdG}(\mathbf{r}) &=
\left(\begin{matrix}
\hat{H}(\mathbf{r}) & \Delta_{eff}(\mathbf{r})\\
\Delta_{eff}(\mathbf{r})^* & -\hat{H}(\mathbf{r})^*
\end{matrix}\right), \\
\hat{H}(\mathbf{r}) &= -\nabla^2 + V_{eff}(\mathbf{r}) - \mu.
\end{align}
Here, $\mu$ is the chemical potential, $z = \epsilon + i\delta$ and the positive limit is taken such that $\delta \rightarrow 0^+$. The impurity system is solved via a Dyson equation, 
\begin{equation}
\hat{G}^{imp}_{BdG}(z)=\hat{G}_{BdG}(z) +
\hat{G}_{BdG}(z) \left(\begin{matrix} \delta\hat{V} & \delta\hat{\Delta} \\
\delta\hat{\Delta}^\ast & -\delta\hat{V}^\ast
\end{matrix}\right)
\hat{G}^{imp}_{BdG}(z)\ \text{,}
\label{eqn:dyson}
\end{equation}
where the potentials are $ \delta\hat{V} = \hat{V}_{imp} - \hat{V}_{bulk}$ and $ \delta\hat{\Delta} = \hat{\Delta}_{imp} - \hat{\Delta}_{bulk}$.
Here, $\hat{G}_{BdG}(z)$ is the Green's function of the unperturbed but superconducting crystal and $\hat{G}^{imp}_{BdG}(z)$ is the resulting impurity cluster Green's function. The impurity real-space cluster is embedded within the unperturbed superconducting crystal and Eq.~(\ref{eqn:dyson}) is solved self-consistently relaxing the charge and anomalous densities within the finite impurity cluster. 

Within the atomic sphere approximation (ASA) each atom $i$ can be associated with an atomic sphere with the radius $r^{ASA}_i$. Thus the potentials $V_{eff}(\mathbf{r})$ and $\Delta_{eff}(\mathbf{r})$ can be written in sums
\begin{align}
V_{eff}(\mathbf{r}) &= \sum_i V_i(\mathbf{r}), \\
\Delta_{eff}(\mathbf{r}) &= \sum_i \Delta_i(\mathbf{r}),
\end{align}
with $V_i(\mathbf{r}) = 0$ and $\Delta_i(\mathbf{r}) = 0$ if $|\mathbf{r}| \geq r_i^{ASA}$ and Equ.~(\ref{eqn:Deltaeff}) becomes
\begin{equation}
\Delta_i(\mathbf{r}) = \Lambda_i\chi_i(\mathbf{r}).
\end{equation}


\section{\label{sec:3} Niobium with Impurities}

\begin{figure}[h]
\centering
\includegraphics*[width=1\linewidth,clip]{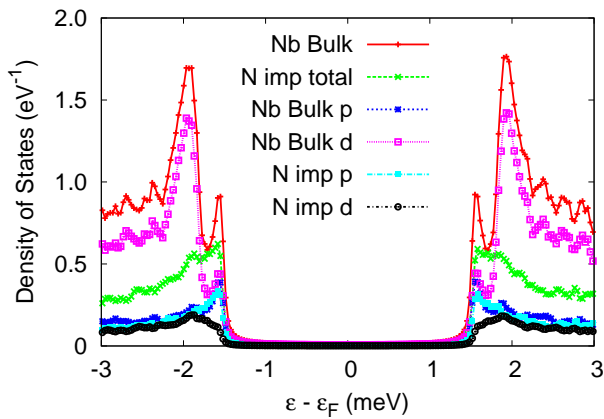}
\caption{LDOS (s,p,d, and total) of an N impurity in comparison to the DOS of unperturbed periodic Nb.}
\label{fig:Nimp_spd}
\end{figure}

As a first test we consider N impurities in Nb, a conventional impurity in this elemental superconductor. On one hand we aim to analyse the effect of the impurity on the superconducting state in the surrounding Nb. On the other hand we will explore the interplay between the gap anisotropy as discussed in detail in Ref.~\onlinecite{Saunderson2020} and the electron scattering off the substitutional impurity. The self-consistent impurity cluster contains 89 atoms where the boundary condition is the perfect superconducting periodic crystal. The central atom is replaced by a substitutional N impurity, the interaction parameter at the impurity site is $\Lambda_{imp}=0$ and we relax the normal charge density, $\rho({\bf r})$, as well as the anomalous density, $\chi({\bf r})$, within the impurity cluster. The LDOS at the central N impurity is shown in Fig.~(\ref{fig:Nimp_spd}) and compared to the Nb DOS of the periodic superconductor. As we set $\Lambda_{imp}=0$, the quasiparticle gap at the N site is purely induced from the surrounding superconducting Nb. As such it has the same principle size as Nb with a notable absence of the outer coherence peak. This follows from the lack of d-states in N and the fact that only the inner coherence peak of Nb has a significant p-character. The larger gap, outer coherence peak, in Nb is almost entirely of d-character.

\begin{figure}[h]
\centering
\includegraphics*[width=1\linewidth,clip]{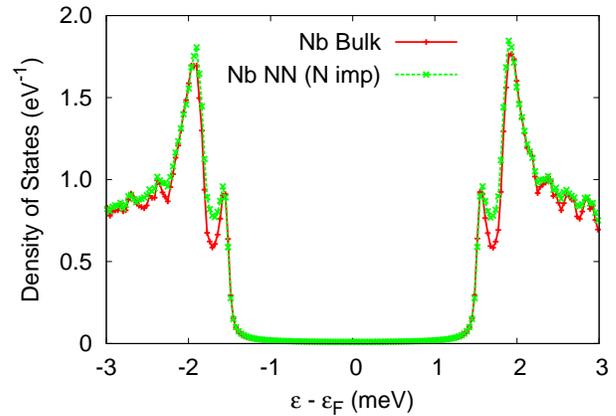}
\caption{The DOS of unperturbed periodic Nb in comparison to the LDOS of the nearest neighbour Nb atom atom in the impurity cluster next to N.}
\label{fig:Nimpscattering}
\end{figure}

Reversing this argument the N impurity should induce strong scattering for the d-electrons of the surrounding Nb. This effect is highlighted in Fig.~(\ref{fig:Nimpscattering}) where the LDOS of the nearest neighbour Nb atom adjacent to the N impurity is displayed. A clear broadening between the inner and outer coherence peak is visible, indicating the challenge to resolve the gap anisotropy in this elemental superconductor when structural or chemical perturbations are present. As in real materials such perturbations will be inevitable, making it demanding experimentally to clearly resolve gap anisotropies on the relevant energy scales.

\begin{figure}[h]
\centering
\includegraphics*[width=1\linewidth,clip]{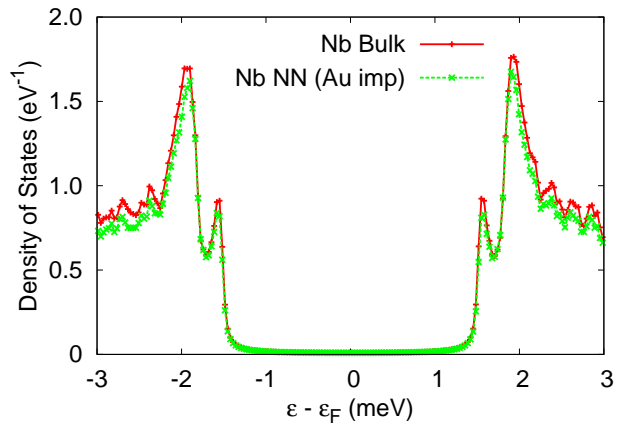}
\caption{DOS of unperturbed periodic Nb in comparison to the LDOS of a nearest neighbour Nb atom in the impurity cluster next to the Au impurity.}
\label{fig:Auimpscattering}
\end{figure}

In order to support our argument that the lack of d orbital character at the N impurity site is responsible for the effective broadening of the peaks we compare the previous result to a situation where the substitutional impurity is Au contributing significant d-character. Figure~\ref{fig:Auimpscattering} clearly shows the lack of broadening between the inner and outer coherence peak as the the DOS of the unperturbed periodic Nb is compared to the LDOS of the nearest neighbour Nb in the impurity cluster containing Au at its centre.

\section{\label{sec:4}Granular Superconductivity}

So far we have analysed the induced superconductivity at the impurity site as well as the impact of the electronic scattering by the impurity atom on the surrounding superconductor. The fact that the superconductor induces a superconducting gap at the impurity site without an effective interaction parameter is not new in principle and has been investigated before \cite{Fetter1965,Flatte1997}. In the following we will investigate the inverse problem, where a superconducting impurity cluster is embedded in a non-superconducting material. The relevance of this granular superconductivity is its connection to the pseudo gap phase of underdoped high-Tc cuprate superconductors. In general, superconductivity emerges from two distinct quantum phenomena: pairing between electrons and long range phase coherence. In conventional BCS theory, the condensation of Cooper pairs into a phase-coherent, quantum state takes place simultaneously at the transition temperature. However, in the underdoped high-Tc cuprate superconductors the electron pairing occurs at higher temperatures than the long-range phase coherence \cite{Emery_1995}. In addition, this as been observed in some disordered, amorphous, superconductors \cite{Dubouchet_2019}. In this model of granular superconductivity the existence of preformed Cooper pairs, pairing without long range phase coherence, are showing similarities with the pseudogap regime of underdoped high-Tc cuprate superconductors.

For a relatively small cluster of material with non-zero interaction parameter embedded in a normal metal the superconductivity will be suppressed and the quasiparticle gap will be forced to close. However, if such a cluster reaches the size of the corresponding superconducting coherence length, $\xi_0$, the expectation is that superconductivity can be sustained within the cluster. Within BCS theory \cite{Ketterson1999a} the coherence length is given by\begin{equation}
\label{eqn:CoherenceLengthBCS}
\xi_0 = \frac{\hbar \nu_F}{\pi\Delta},
\end{equation}
where $\nu_F$ is the Fermi velocity, linking the coherence length to the inverse size of the superconducting gap $\Delta$.~\footnote{Although we use a local pairing model it leads to Cooper pairs which are extended in real space since Cooper pairs are formed in momentum space and not in real space. In fact our local pairing model is the analogue of conventional BCS theory, where Cooper pairs are formed by electrons with different quantum numbers $\{\mathbf{k}, \protect\uparrow\}$ and $\{ -\mathbf{k},\protect\downarrow\}$ in which states from the region of the gap around the Fermi level are mixed. The coherence length is the extension of these wave packets in real space which is only indirectly related to the pairing model.} 
The coherence length of bulk Nb is approximately $38$nm \cite{Kittel2005imp}. A cluster of that size would roughly contain $10^6$ atoms and is beyond any capability of our method. Within our standard calculations cluster of a few hundred atoms could be considered, limiting the cluster size to $<2nm$. However, it is still possible to test the relation $\xi_0\sim\Delta^{-1}$ for artificially enlarged superconducting gaps.

\begin{figure}[h]
\centering
\includegraphics*[width=1\linewidth,clip]{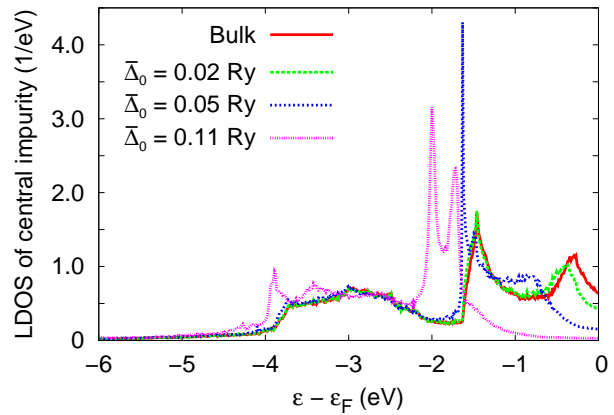}
\caption{The LDOS of the central atom in a cluster of 89 atoms with a constant $\Delta_{eff}$ for every atom in the cluster embedded in metallic Nb with $\bar{\Delta}_0$ defined in Eq.~(\ref{Eq.:Def_Deltabar}). }
\label{fig:FlatGapOneShot}
\end{figure}

The cluster was constructed from Niobium atoms with a non-zero interaction parameter $\Lambda_i$, embedded in an infinite normal state Nb crystal. In a first step we omit self-consistency and explore the resulting superconducting gap in LDOS calculations when a constant pairing potential $\Delta_{eff}(\mathbf{r}) = \Delta_{eff}$ is applied. Figure~\ref{fig:FlatGapOneShot} shows the LDOS of the central atom as we change the constant $\Delta_{eff}$. For a pairing potential of $\Delta_{eff} = 0.11 Ry$  a gap in the quasiparticle spectrum of approximately $\Delta = 1.5\mathrm{eV}$ is induced, decreasing in size quickly with the size of the pairing potential. At $\Delta_{eff}=5\times10^{-2}\mathrm{Ry}$ a suppression of the LDOS is still visible without a full opening of a gap and at $\Delta_{eff}=2\times10^{-2}\mathrm{Ry}$ only a small deviation from the non-superconducting Nb remains. This implies that the surrounding metallic Nb enforces a suppression of the superconducting state as soon as the pairing potential is smaller than $5\times10^{-2}\mathrm{Ry}$.

\begin{figure}[h]
\centering
\includegraphics*[width=1\linewidth,clip]{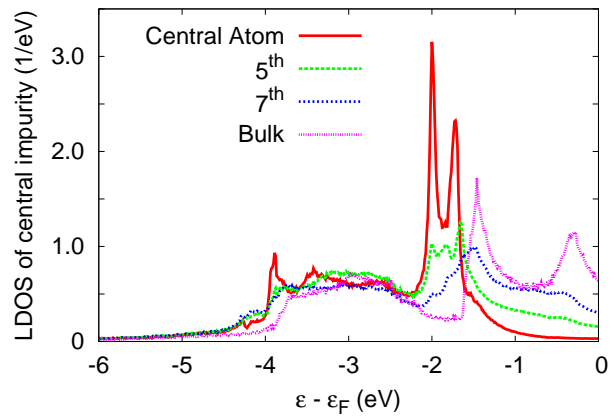}
\caption{The LDOS of atoms in the n$^{\mathrm{th}}$ shell in a cluster of 89 atoms with a constant effective pairing potential $\Delta_{eff} = 0.11\mathrm{Ry}$, for every atom in the cluster.}
\label{fig:FlatGapFurtherOut}
\end{figure}

In a next step we fix the pairing potential at $\Delta_{eff} = 0.11~\mathrm{Ry}$ and explore how the gap in the LDOS develops as we are approaching the boundary to the metallic Nb. In Fig.~\ref{fig:FlatGapFurtherOut} the corresponding results are summarized, comparing the central atom to the 5th nearest ($0.57$~nm) and the 7th nearest ($0.72$~nm) shell. Even at a distance of $0.57$~nm the coherence peak is still visible but the original gap is fully filled with a slightly suppressed LDOS and the local gap is gradually disappearing. There is no sudden transition from a gapped to a normal state implying the coexistence of anomalous (pairing) as well normal (electron) density.

\begin{figure}[h]
\centering
\includegraphics*[width=1\linewidth,clip]{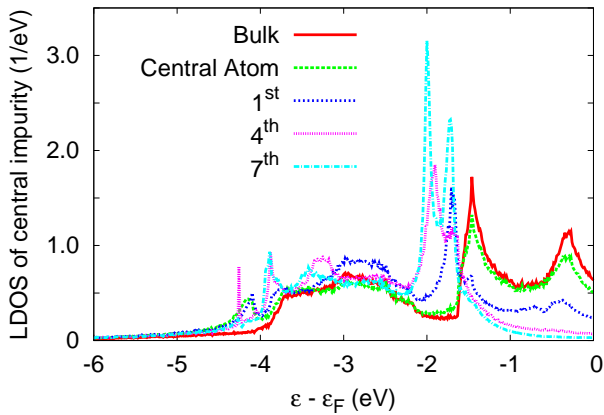}
\caption{The LDOS of the central atom in the cluster of 89 atoms with a constant pairing potential $\Delta_{eff}=0.11\mathrm{Ry}$ for every atom up to and including the atoms in the n$^{\mathrm{th}}$ shell.}
\label{fig:FlatGapNearestNeighbour}
\end{figure}

This finding is very similar to the situation where we change the size of the region within the cluster for which we consider a non-zero and constant pairing potential. The resulting LDOS for the central atom is shown in Fig.~\ref{fig:FlatGapNearestNeighbour}. Again the coherence peak is more or less visible down to a region of nearest neighbours only but the LDOS at the Fermi energy increases as the cluster is decreasing.

\begin{figure}[h]
\centering
\includegraphics*[width=1\linewidth,clip]{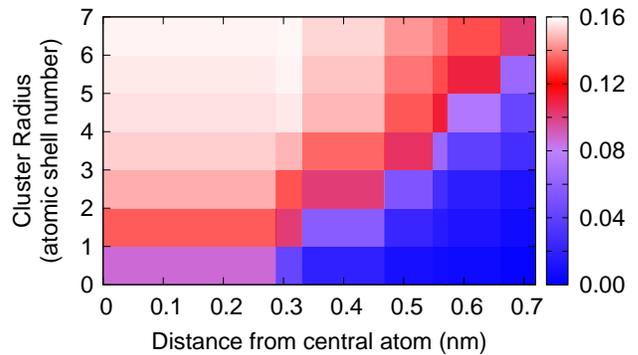}
\caption{The average anomalous charge $\bar{\chi}_i$ per atom as a function of distance in the cluster for the one-shot gap calculations corresponding to Fig.~\ref{fig:FlatGapNearestNeighbour}.}
\label{fig:NearestNeighbourAnomRho}
\end{figure}

To summarize those findings we define the anomalous charge $\bar{\chi}_i$,
\begin{equation}
\bar{\chi}_i = \frac{1}{V_{WS}}\int_0^{r_{ASA}}d^3r\chi_i(\mathbf{r}),
\end{equation}
which is a constant for each shell at a given distance from the central atom within the cluster. Correspondingly, we define the average gap $\bar{\Delta}$,
\begin{equation}\label{Eq.:Def_Deltabar}
\bar{\Delta}_i = \frac{1}{V_{WS}}\int_0^{r_{ASA}}d^3r\Delta_{i}(\mathbf{r})
\end{equation}
which in the self-consistent calculations is related to $\bar{\chi}_i$ by the proportionality $\Lambda_i$. However, in non-self consistent one shot calculations the relation is more complex. In Fig.~\ref{fig:NearestNeighbourAnomRho} we summarize the results for the anomalous charge $\bar{\chi}_i$ as we change the region of non-zero $\bar{\Delta}_i$ (y-axis in Fig.~\ref{fig:NearestNeighbourAnomRho}) corresponding to Fig.~\ref{fig:FlatGapNearestNeighbour}, while at the same time analysing the full cluster (x-axis in Fig.~\ref{fig:NearestNeighbourAnomRho}). In all cases the anomalous charge is quickly reduced if we consider atoms outside the region of the applied non-zero $\bar{\Delta}_i$. Nevertheless, it is clearly visible how the anomalous charge is enhanced at the central atom as the region of non-zero $\bar{\Delta}_i$ is increased, while at the same time a small anomalous charge is induced beyond the region of non-zero pairing potential.

\begin{figure}[h]
\centering
\includegraphics*[width=1\linewidth,clip]{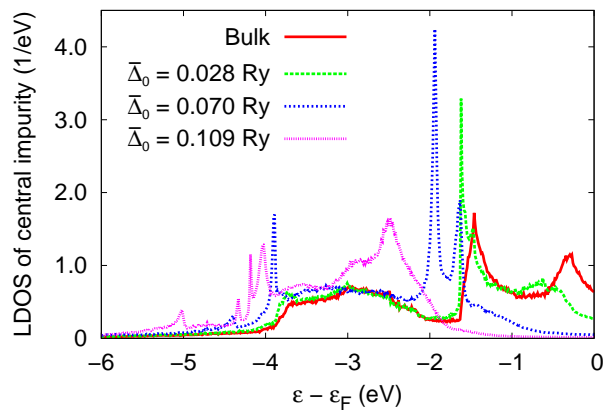}
\caption{The LDOS of the central atom in the cluster of 89 atoms with a constant $\Lambda_i$ applied to every atom in the cluster. The legend specifies the size of $\bar{\Delta}_0$ of the central atom. For the green line $\Lambda = 0.3 Ry$, blue $\Lambda = 0.4 Ry$, pink $\Lambda = 0.5 Ry$.}
\label{fig:SCFimpuritycluster}
\end{figure}

In order to make a direct connection to the coherence length and its relation to the superconducting gap it is important to perform all calculations self-consistently. According to the BCS result the cluster needs to be larger than the coherence length to support superconductivity. The complication arises from the fact that we observe pairing (anomalous charge in Fig.~\ref{fig:NearestNeighbourAnomRho}) while no gap is induced in the quasiparticle spectrum (LDOS, see Fig.~\ref{fig:FlatGapNearestNeighbour}). In Fig.~\ref{fig:SCFimpuritycluster} we summarize the fully self-consistent calculations changing the constant interaction $\Lambda_i$ for the full cluster of 89 atoms. Shown is the LDOS of the central atom. Similarly to our discussion before, the gap in the LDOS vanishes as we reduce the interaction parameter to $\Lambda_i=0.3Ry$ while the corresponding average gap and thus the pairing potential stays non-zero as highlighted in the legend.

\begin{figure}[h]
\centering
\includegraphics*[width=1\linewidth,clip]{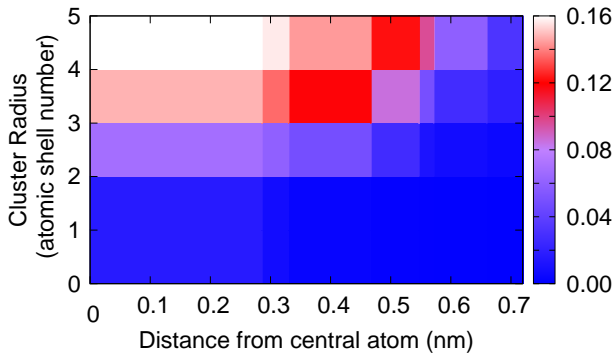}
\caption{The average anomalous charge $\bar{\chi}_i$ per atom as a function of distance in the cluster. This is the corresponding figure to Fig.~\ref{fig:FlatGapNearestNeighbour} but for a fully self-consistent calculation.}
\label{fig:NearestNeighbourAnomRho_self}
\end{figure}

The equivalent summary for the self-consistent calculations to Fig.~\ref{fig:NearestNeighbourAnomRho} in case of the one-shot is shown in Fig.~\ref{fig:NearestNeighbourAnomRho_self}. A much sharper transition between a vanishing gap is visible in case of a cluster with an applied interaction up to the second shell only.

\begin{figure}[h]
\centering
\includegraphics*[width=1\linewidth,clip]{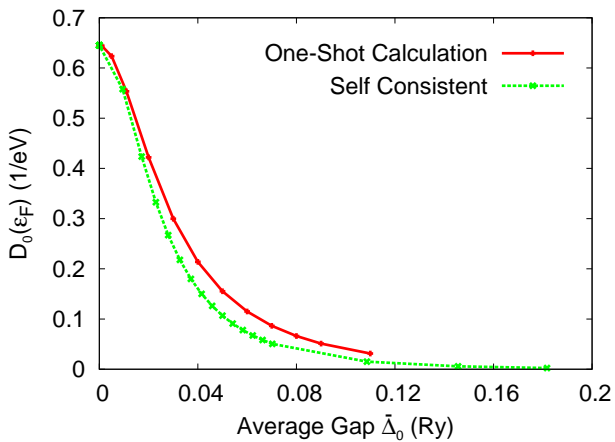}
\caption{The LDOS at the Fermi level as a function of the average gap $\bar{\Delta}_0$ for the central atom.}
\label{fig:Gap_VS_DOS}
\end{figure}

In order to better understand the relationship between self-consistency and one-shot LDOS calculations, we have to analyse the relationships between the anomalous charge, $\bar{\chi}_i$, the average gap, $\bar{\Delta}_i$ and the LDOS at $\epsilon_F$, $D_i(\epsilon_F)$. In Fig.~\ref{fig:Gap_VS_DOS} $D_0(\epsilon_F)$ as a function of $\bar{\Delta}_0$ is shown for the central atom in a cluster of 89 atoms. For the self consistent calculation, a non-zero and constant $\Lambda_i$ is applied to all of the atoms up to the $7$th nearest neighbour shell. For the one-shot calculations a constant pairing potential with the corresponding average gap is applied to all atoms. In this representation both approaches give very similar results. In all these cases we observe a smooth transition between the opening of a gap in the quasiparticle spectrum gradually closing in as we change the superconducting strength, either via the interaction parameter, the average applied gap or the size of the superconducting region.

\begin{figure}[h]
\centering
\includegraphics*[width=1\linewidth,clip]{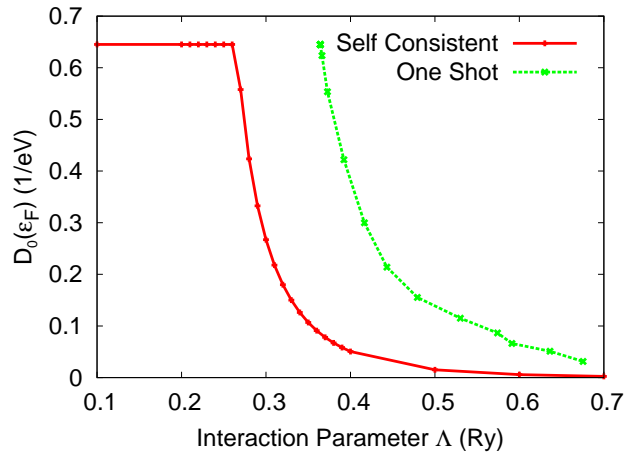}
\caption{The LDOS for the central atom at the Fermi level as a function of the interaction parameter. In case of the one-shot calculations, the ratio between the applied average pairing potential and the anomalous charge is used to define an effective interaction parameter. }
\label{fig:Int_VS_Def}
\end{figure}

However, according to BCS theory, there should be a sharp transition where a gap is induced once the coherence length is reached. This sharp transition becomes visible as we show $D_0(\epsilon_F)$ as a function of $\Lambda_i$ applied to a cluster of 89 atoms in Fig.~\ref{fig:Int_VS_Def}. This representation highlights the differences between the one shot and the fully self-consistent calculations. Only for the self-consistent calculations we are a able to observe the sharp transition at which the system becomes superconducting at a non-zero interaction parameter. For the one-shot calculations as we reduce the applied average gap we will in all cases observe a non-zero induced anomalous density and as such an effective interaction parameter. In contrast for the self-consistent calculation as we reduce the interaction parameter we eventually reach the point where all superconductivity is suppressed, the anomalous density goes to zero, the gap in the quasiparticle spectrum vanishes and we observe a phase transition. However, while at this point Cooper pairs start to form we do not observe full phase coherence why we do not observe the opening of a gap in the LDOS as discussed earlier.

\begin{figure}[h]
\centering
\includegraphics*[width=1\linewidth,clip]{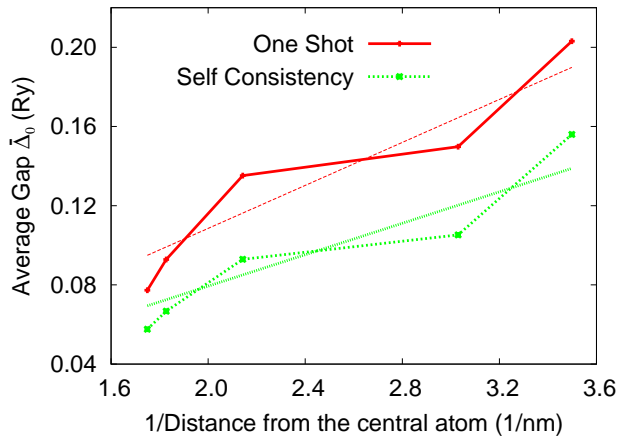}
\caption{The average gap $\bar{\Delta}$ as a function of the inverse radius of the region of superconducting atoms.}
\label{fig:Gap_VS_ClusRad}
\end{figure}

To finally investigate the coherence length within our method we show in Fig.~\ref{fig:Gap_VS_ClusRad} the average gap as a function of the size of the region with a non-zero interaction parameter. In order to generate this figure we consider both cases the one shot and the the self consistent calculations. In case of the one shot calculations at a given cluster size we increase the average gap, $\bar{\Delta}_i$, until the DOS at the Fermi energy is suppressed below $0.1(eV)^{-1}$. For the self-consistent calculations we do the same but varying the interaction parameter ($\Lambda_i$) until we reach the same threshold. The chosen threshold is a trade off between numerical accuracy and reaching a fully gapped situation. According to the BCS relation, Eq.~(\ref{eqn:CoherenceLengthBCS}), we should find $\bar{\Delta}\propto 1/\xi_0$ with the slope given by $\frac{\hbar v_F}{\pi}$. We find a roughly linear dependence with the linear fit giving the slope and as such the Fermi velocity to be $3.52\times10^{6}m/s$ and $2.58\times10^{6}m/s$ for the one shot and the self-consistent calculations, respectively. This is in reasonable agreement to typical Fermi velocities of the order of $1\times10^{6}m/s$.

\section{\label{sec:5}1D Chain Results}

In the previous section, the formation of the full gap in the DOS as a function of the cluster size is somewhat complicated by the limited number of atoms which can be treated in a fully {\it ab initio} calculation.  Here, we performed a very similar calculation, but on a much simpler system, the 1D chain of periodic, uniform square well potentials. In this model, a finite chain of square well potentials is embedded into an infinite chain of slightly different square well potentials. A certain advantage of the KKR and BdG-KKR theory is, that it can be formulated in a formally similar way to its 3D counterpart described in section \ref{sec:2} \cite{Suvasini1992,Butler1976a}. 
It has the advantage that it can be solved for cluster sizes which are practically impossible in a 3D calculation. To mimic more closely the 3D system in the superconducting case, we set the effective pair interaction to zero in the infinite chain and to a finite value in the impurity region. Then the equations are solved numerically for the embedded cluster of various sizes.

\begin{figure}[h]
\centering
\includegraphics*[width=1\linewidth,clip]{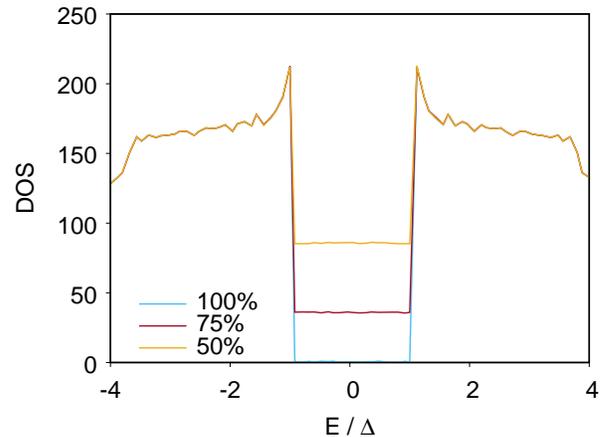}
\caption{The local Density of sates (DOS) of the central atom as the chain of superconducting atoms embedded in a metallic systems is increased
in length. $100\%$ refers to the length where there is a full gap. Accordingly shorter lengths are normalized to this.}
\label{fig:1dchain_DOS}
\end{figure}

\begin{figure}[h]
\centering
\includegraphics*[width=1\linewidth,clip]{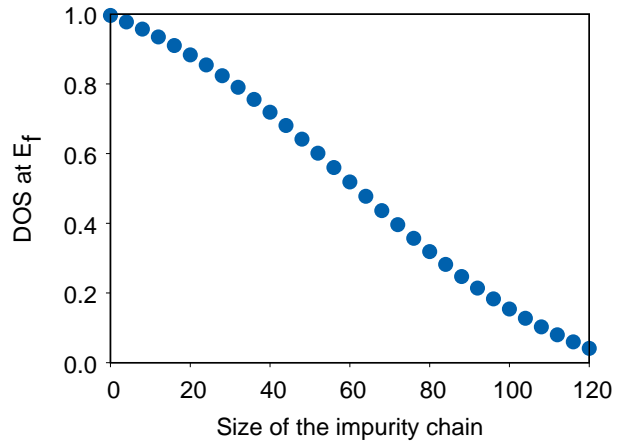}
\caption{
The LDOS at the Fermi level for the central atom as a function of the chain length (number of square well potentials).}
\label{fig:1dchain_DOScomp}
\end{figure}

The analogue to Fig.~\ref{fig:FlatGapNearestNeighbour} is Fig.~\ref{fig:1dchain_DOS}, considering the LDOS around the Fermi energy of the central atom within the cluster. Evidently, the quasiparticle spectrum does not exhibit a full gap until the size of the cluster reaches a critical length, the coherence length. This is in full analogy to the 3D system shown in Fig.~\ref{fig:FlatGapNearestNeighbour}. In the same way as before the LDOS is suppressed around $E_F$ for all cluster sizes while the coherence peaks stay rather constant. In this simple 1D model the width is equal to the applied effective pair interaction. To further illustrate this behaviour, Fig.~\ref{fig:1dchain_DOScomp} shows how the DOS at the Fermi level of the central atom (square well potential) behaves as the size of the impurity chain increases.

\begin{figure}[h]
\centering
\includegraphics*[width=1\linewidth,clip]{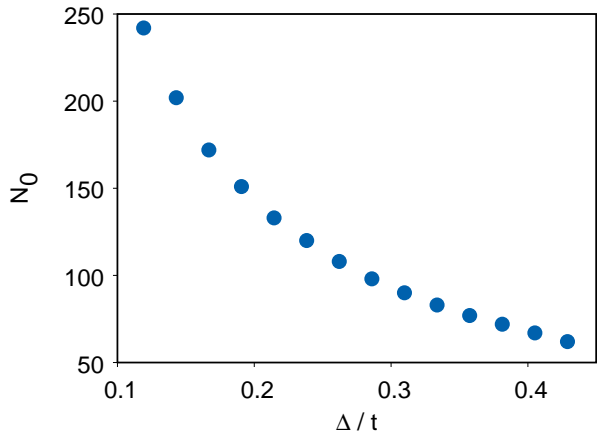}
\caption{The minimum cluster size (N$_0$) at which a gap can be observed as a function of the effective pair interaction in the BdG equations. In this 1d model the effective pair interaction is shown in the units of the single band width $t$.}
\label{fig:Cluster_size_vs_gap}
\end{figure}

Fixing the gap and extracting the chain length for which the LDOS at $E_f$ vanishes and repeating this calculation for a range of gaps we summarize these results in Fig.~\ref{fig:Cluster_size_vs_gap}. As for the 3D case this relation should be 
be compared to Eq.~\ref{eqn:CoherenceLengthBCS}. Here the relationship between the gap and the length of the 1D impurity cluster almost perfectly fulfils the prediction of BCS theory. We again may conclude, that the minimum cluster size with a true superconductivity gap is the coherence length.

\section{Discussion and Conclusions}

We have implemented a self-consistent solution of the BdG equations into a real-space impurity solver within the KKR formalism, extending the formalism from our previous work \cite{Saunderson2020}. In this formalism both charge, $\rho(\mathbf{r})$ and anomalous $\chi(\mathbf{r})$ densities can be relaxed, with $\Lambda$ being the interaction parameter which drives the superconductivity.

From our previous work we showed that the gap anisotropy in Nb is successfully reproduced. Here we show that in the presence of impurities, that gap anisotropy gets broadened by impurity scatterers which contain no `d' states as that is the main contributor to density of states around the gap. It was not possible to find an impurity which only contained s orbitals at this energy level, potentially obscuring the peaks entirely. However, introducing impurity scattering from Au, an element with `p' and `d' character close to the Fermi level no broadening of the peaks at all was observed. This confirmed our argument and underlined the importance of the detailed knowledge of the orbital character of the impurity electrons.

After this we inverted the problem, considering the effect of a non-superconducting bulk on a cluster of superconducting impurity atoms. We found that the bulk strongly influences the impurity atoms, similar to our previous study. The gap and the corresponding interaction parameter had to be artificially increased by approximately 1000 times in order to induce a gap within an 89 atom cluster. This is directly related to the superconducting coherence length of the superconducting material. We showed that we were able to reproduce the BCS expression of the coherence length as a function of the superconducting gap. However, we would like to highlight that we clearly observe distinct states of our system. Below a certain threshold the interaction is too weak and superconductivity is suppressed throughout the entire system. Passing a critical value we observe the formation of Cooper pairs without full phase coherence leading to a suppression of the LDOS at the Fermi energy without the formation of a full energy gap. Only upon increasing the interaction parameter further phase coherence across the system is achieved and a full gap opens.

To solidify this claim, we perform a simplified 1d chain KKR model. Here we showed that the coherence length obeys the same trend as for the 3d KKR method. However, due to the easier numerical implementation much larger systems could be explored displaying the relation in a much clearer way. 

In summary we have showed that using a fully \textit{ab initio} method to describe the normal state and a simple phenomenological parametrisation to describe the superconducting exchange correlation functional we can describe the effect of impurities on the superconducting state. Even in the presence of impurities it is still possible to observe the gap anisotropy in Nb while depending on the orbital character of the impurity atom a significant broadening of the coherence peaks can be observed. In addition, we have applied a direct method to test the coherence length of superconducting materials which is concurrent with BCS theory. Our future aim will be to include magnetism and spin-orbit coupling to look at more exotic phenomena associated with impurities including Yu-Shiba-Rusinov states and the generation of triplet currents.

\section{Acknowledgements}

This work was carried out using the computational facilities of the Advanced Computing Research Centre, University of Bristol - http://www.bris.ac.uk/acrc/ .
It was financially supported by the Centre for Doctoral Training in Condensed Matter Physics, funded by EPSRC EP/L015544/1. B. Újfalussy was supported by the Hungarian National Research, Development and Innovation Office under contract OTKA K115632 and the BME Nanotechnology FIKP grant (BME FIKP-NAT). G. Csire gratefully acknowledges support from the European Union’s Horizon 2020 research and innovation programme under the Marie-Sklodowska-Curie grant agree- ment No. 754510. The authors would like to thank M. Czerner and Prof Heiliger for fruitful discussions on the KKR method . In addition, thanks to Ming-Hung Wu and Reena Gupta for many helpful discussions related to the project.

\bibliography{References}
\bibliographystyle{apsrev4-2}

\end{document}